\documentclass[twocolumn,prl,aps,superscriptaddress]{revtex4-1}
\usepackage[latin9]{inputenc}
\usepackage{mathtools}
\usepackage{amsbsy}
\usepackage{amstext}
\usepackage{graphicx}% Include figure files
\usepackage[unicode=true,pdfusetitle,
 bookmarks=false,
 breaklinks=false,pdfborder={0 0 1},backref=false,colorlinks=false]
 {hyperref}
 \usepackage{float}
 \usepackage[normalem]{ulem}
\hypersetup{
 bookmarksnumbered=false,bookmarksopen=false}
\makeatletter

\@ifundefined{textcolor}{}
{%
 \definecolor{BLACK}{gray}{0}
 \definecolor{WHITE}{gray}{1}
 \definecolor{RED}{rgb}{1,0,0}
 \definecolor{GREEN}{rgb}{0,0.7,0}
 \definecolor{BLUE}{rgb}{0,0,1}
 \definecolor{CYAN}{cmyk}{1,0,0,0}
 \definecolor{MAGENTA}{cmyk}{0,1,0,0}
 \definecolor{YELLOW}{cmyk}{0,0,1,0}
}

\newcommand{\beq}{\begin{equation}}
\newcommand{\eeq}{\end{equation}}
\newcommand{\beqa}{\begin{eqnarray}}
\newcommand{\eeqa}{\end{eqnarray}}

\usepackage{amsmath}
\usepackage{graphicx}
\usepackage{amssymb}
\usepackage{txfonts,color}
\makeatother

\begin{document}
\title{High-Resolution NMR Spectroscopy at Large Fields with Nitrogen Vacancy Centers}
\author{C. Munuera-Javaloy}
\affiliation{Department of Physical Chemistry, University of the Basque Country UPV/EHU, Apartado 644, 48080 Bilbao, Spain}
\affiliation{EHU Quantum Center, University of the Basque Country UPV/EHU, Leioa, Spain}
\author{A. Tobalina}
\affiliation{Department of Physical Chemistry, University of the Basque Country UPV/EHU, Apartado 644, 48080 Bilbao, Spain}
\affiliation{EHU Quantum Center, University of the Basque Country UPV/EHU, Leioa, Spain}
\author{J. Casanova}
\affiliation{Department of Physical Chemistry, University of the Basque Country UPV/EHU, Apartado 644, 48080 Bilbao, Spain}
\affiliation{EHU Quantum Center, University of the Basque Country UPV/EHU, Leioa, Spain}
\affiliation{IKERBASQUE, Basque Foundation for Science, Plaza Euskadi 5, 48009 Bilbao, Spain}

\begin{abstract}
Ensembles of nitrogen-vacancy (NV) centers are used as sensors to detect NMR signals from micron-sized samples at room temperature. In this scenario, the regime of large magnetic fields  is especially interesting as it leads to a large nuclear thermal polarisation --thus, to a strong sensor response even in low concentration samples-- while chemical shifts and J-couplings become more  accessible. Nevertheless, this  regime remains largely unexplored owing to the difficulties to couple NV-based sensors with high-frequency nuclear signals. In this work, we circumvent this problem with a method that maps the relevant energy shifts in the amplitude of an induced nuclear spin signal that is subsequently transferred to the sensor. This stage is interspersed with free-precession periods of the sample nuclear spins where the sensor does not participate. Thus, our method leads to high spectral resolutions ultimately limited by the coherence of the nuclear spin signal.
\end{abstract}
\maketitle

\emph{Introduction.-} 
Nuclear magnetic resonance (NMR) is a fundamental technique for a variety of  areas such as diagnosis medicine, biochemistry, and analytical chemistry~\cite{Abragam61, Levitt08}. From its inception in 1946~\cite{Purcell46, Bloch46, PhysNobel52} NMR has grown to constitute its own research area~\cite{Becker93}. The discussion over the optimal field strength has been always present in the field~\cite{Hoult86}, and the profits provided by elevated magnetic fields of the order of several Teslas--namely, increased spatial and spectral resolutions-- have long been known~\cite{Ugurbil03,Moser17}. In recent years NMR has experienced a profitable simbiosis with the rapidly growing field of quantum technologies~\cite{Dowling03}. In particular, the use of newly-developed solid-state quantum sensors~\cite{Degen17}, such as NV centers in diamond~\cite{Doherty13}, has enabled to interrogate ever smaller samples~\cite{Balasubramanian08,Maletinsky12,Staudacher13}. This has led to NMR experiments with unprecedented spatial resolutions, even reaching single molecules~\cite{Shi15,Lovchinskya2016,Munuera2021a}. In this regard, the benefits of operating at large magnetic fields are expected to carry on for NMR analysis of micro and nanoscale sized samples with quantum sensors.

Nuclear spins are main actors in NMR as they are the source of the target magnetic signal. The evolution of a nucleus in an external magnetic field is also affected by the distribution of nearby magnetic sources such as electrons in chemical bounds. Consequently, detecting the resulting variations in the Larmor precession frequency through NMR procedures --thus, leading to precise information about J-couplings and chemical shifts-- serves as an accurate diagnosis of the molecular structure around target nuclei. Identifying these changes requires measurement protocols that achieve frequency resolutions of the order of Hz. However, the spectral resolution of standard quantum sensing techniques is severely limited by the coherence time of the sensor. In the case of NV centers, this restriction leads to kHz resolutions even when the sensor is stabilised with dynamical decoupling techniques~\cite{Munuera2021b}, leading to an insufficient record for useful chemical analysis. 

\begin{figure}[t]
%\begin{center}
\includegraphics[width= 1.\linewidth]{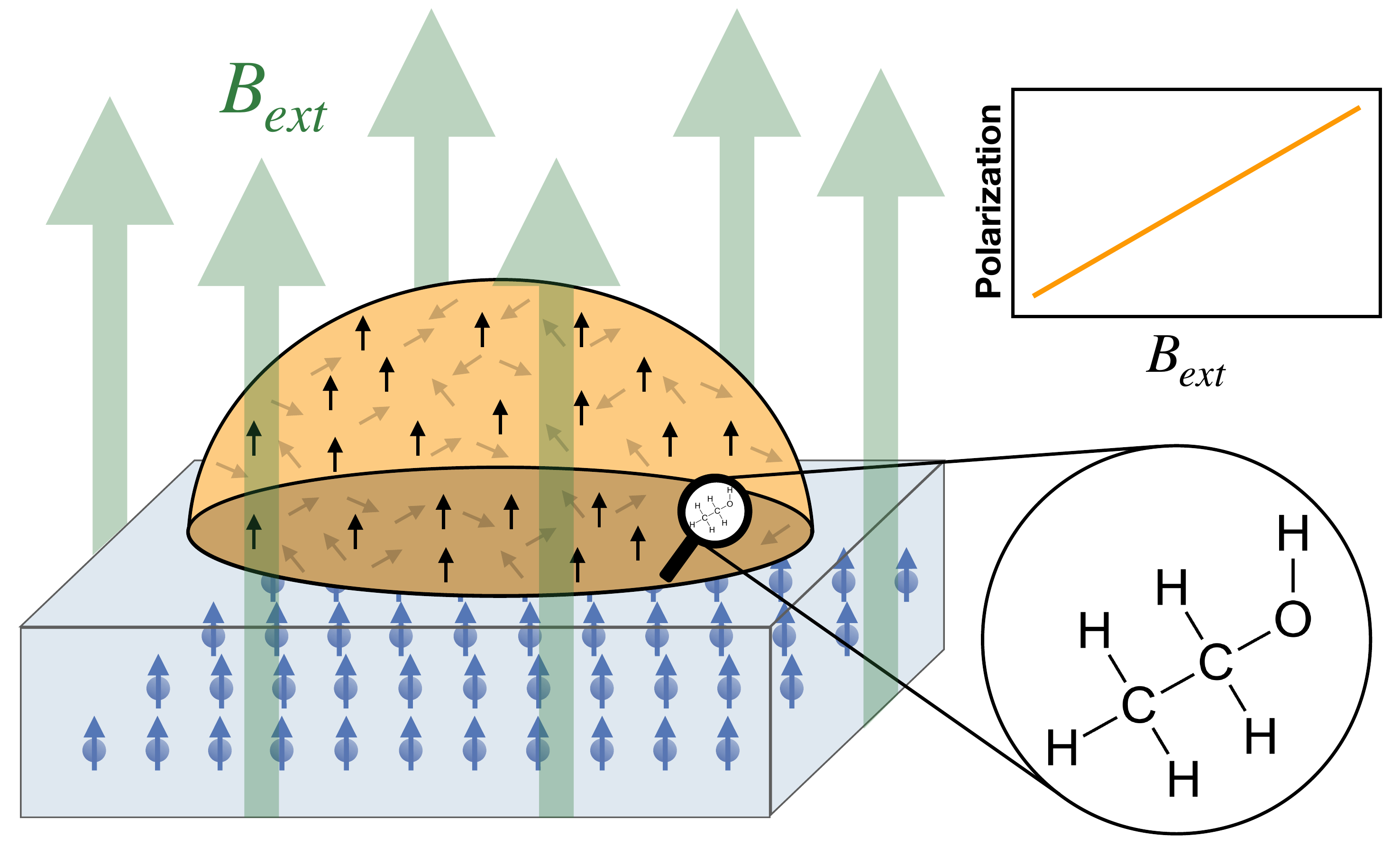}
\caption{\label{fig1} The setup consists on a picoliter  sample placed on a diamond. This contains an NV ensemble as the sensor of the magnetic signal generated by the analyte. The protons of the studied molecule emit a signal that depends on their local environment, thus carrying structural information.}
%\end{center}
\end{figure}

Recently, protocols capable of overcoming these limitations have been devised. With techniques that resemble classical heterodyne detection, measurements of artificial signals using NV probes~\cite{Boss17,Schmitt17}  reached $\mu$Hz resolution. In addition, when applied to micron-sized samples these techniques led to the detection of J-couplings and chemical shifts at low magnetic fields~\cite{Glenn18}. These applications suffer from low sensitivity caused by the weakness of the nuclear signals. This imposes the need for a large number of repetitions and/or samples with a large hydrogen concentration such that they provide  sufficiently intense signals, thus limiting their utility for competitive chemical analysis.

\begin{figure*}[t]
%\begin{center}
\includegraphics[width=  \linewidth]{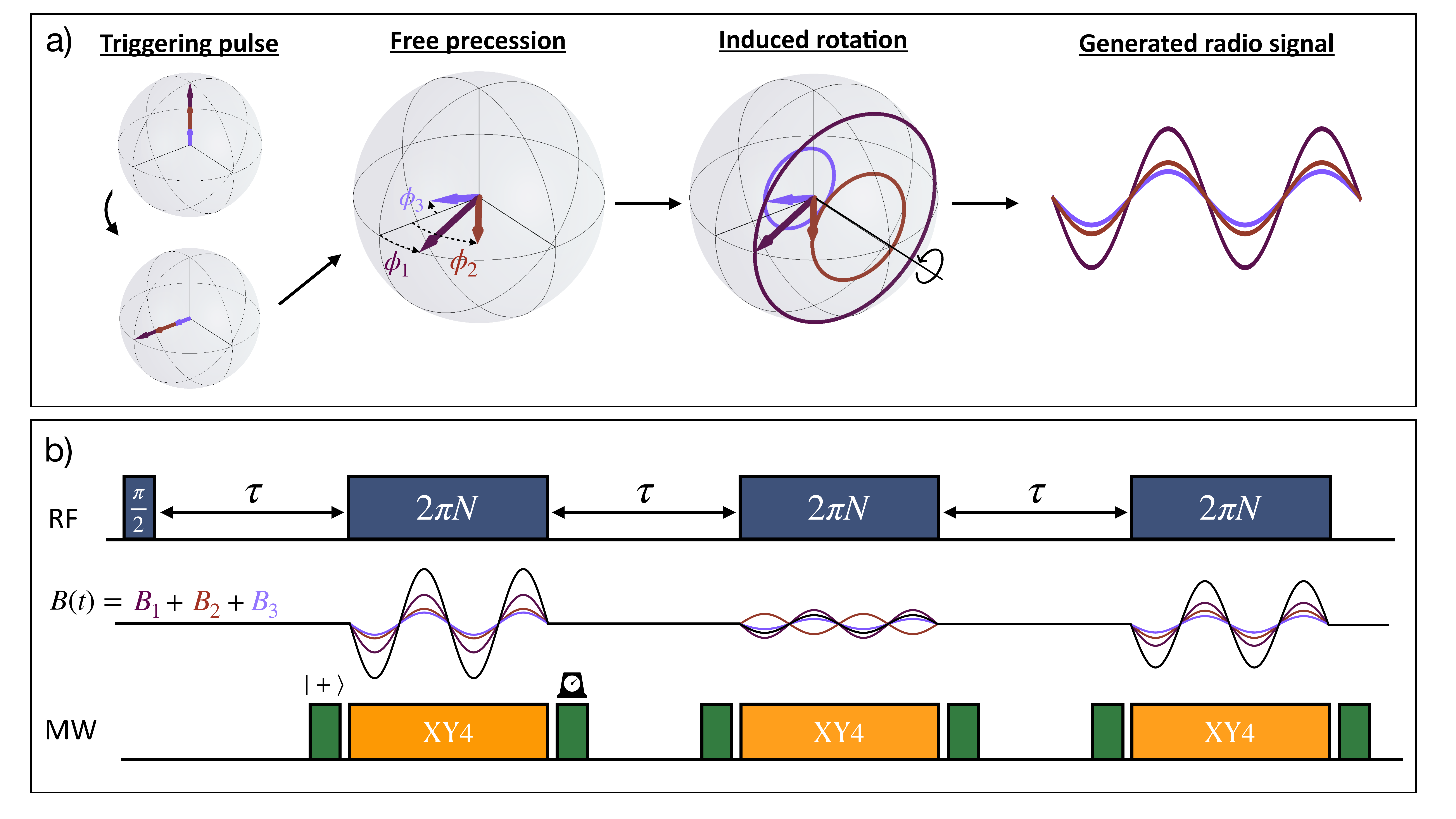} 
\caption{\label{fig2} Custom signal production and measurement. a) An initial RF $\pi/2$ pulse brings the sample thermal polarization to the orthogonal plane and triggers the AERIS protocol consisting on free precessions and induced rotations stages. For a time $\tau$ each magnetization vector $\boldsymbol{M}_k(t)$ precess according to the local field at the position of the nuclear spin. The phase covered by each $\boldsymbol{M}_k(t)$ --this is $\phi_k=\delta_k \tau$-- is encoded in the amplitude of the oscillating field generated via controlled rotations of these vectors. b) (First panel) RF control sequence with interleaved free precessions. (Second panel)  Sample emitted fields. These have different amplitudes due to the distinct projections of each rotating $\boldsymbol{M}_k(t)$ on the Z axis. The depicted case shows three $B_i$ fields as a consequence of the splitting among three magnetization vectors that spin at rates $\delta_1$, $\delta_2$, and $\delta_3$. (Third panel) MW pattern --in our case an XY4 sequence-- on each NV devised to capture the induced signal. Note that the NVs remain inactive during the long free precession stages of the sample, providing our protocol with increased spectral resolution regardless of the sensor coherence time. Prior to the MW sequence, the NV ensemble is initialized in $|+\rangle$ while once its state encodes the desired information it is optically readout in the $\sigma_y$ basis.}
%\end{center}
\end{figure*}

A possible workaround proposes to increment the polarization of the sample using dynamical nuclear polarization techniques, hence achieving improved contrasts~\cite{Bucher20,Arunkumar21}.  Alternatively, operating at large static magnetic fields  enhances the thermal polarisation, increasing the NMR signal intensity without adding new compounds to the sample, hence enabling to interrogate samples in a wide range of concentrations (not only highly concentrated ones). Besides, the presence of large magnetic fields  facilitates the identification of frequency changes caused by the local environment of nuclei, as J-couplings become clearer and chemical shifts increase.

In this Letter we present the AERIS (Amplitude-Encoded Radio Induced Signal) method. This is a detection protocol able to handle large magnetic field scenarios and that achieves a spectral resolution which is only limited by the coherence time of the nuclear spin signal, thus leading to a spectral resolution compatible with chemical shifts and J-couplings. We exemplify the AERIS method with NV centers, however this is equally applicable to other types of solid-state-based sensors~\cite{Soykal2016, Soykal2017}. Moreover, the method might be combined with recently used dynamical nuclear polarization techniques~\cite{Bucher20,Arunkumar21,Maly08,Ni13} leading to stronger target signals. 

\emph{The protocol.-}
State of the art NV-based AC field magnetometry  targets the oscillating signal produced by precessing nuclear spins, whose frequency is proportional to the local field felt by the nuclei. On the one hand, this relation allows to acquire information on the molecular environment of the nuclei by unraveling the spectral composition of the signal. On the other hand, when the sample is exposed to a large magnetic field, it leads to signals that oscillate too fast to be traced by the NV. Note that approaches based on the delivery of appropriately shaped pulses have been proposed for dealing with moderate field scenarios,
or in situations where only reduced MW power is available~\cite{Casanova19, Munuera-Javaloy20}. Here we take an alternative approach and target a deliberately manufactured signal that carries the spectroscopic information of the studied sample encoded in its amplitude rather than in its frequency. 

We consider a thermally polarized sample placed on top of an NV-ensemble-based-sensor and in the presence of a large external magnetic field $B_{ext}$, see Fig.~\eqref{fig1}. The sample would contain a certain type of molecule with nuclear spins in different locations of its structure. Hereafter we use subindex $k$ (or superindex when required by the notation) to
indicate the different precession frequencies produced by distinct local magnetic fields. This scenario is similar to those reported in \cite{Glenn18,Bucher20,Arunkumar21} with the critical difference of the magnitude of $B_{ext}$.

Following~\cite{Levitt08} we describe the spins of our sample via the nuclear magnetization $\boldsymbol M =(M_x,M_y,M_z)$. This is a time-dependent vector proportional to the sample average nuclear magnetic moment. Its behavior during an RF pulse of intensity $\Omega$ in a frame that rotates with the frequency of the RF driving ($\omega$) is described by the Bloch equations
\beq
\label{bloch}
\frac{d}{dt}
\left(\begin{array}{c}
M_x\\
M_y\\
M_z
\end{array}\right)
=
\left(\begin{array}{ccc}
-1/T^*_2 & - \delta & \Omega \sin\phi\\
\delta & -1/T^*_2 & - \Omega \cos\phi\\
-\Omega \sin \phi & \Omega \cos \phi & - 1/T_1
\end{array}\right)
\left(\begin{array}{c}
M_x\\
M_y\\
M_z
\end{array}\right)
+
\left(\begin{array}{c}
0\\
0\\
1/T_1
\end{array}\right),
\eeq
where $\phi$ is the phase of the RF field, and $T_1$ ($T^*_2$) is the nuclear relaxation (dephasing) rate. The detuning $\delta=\omega_L-\omega$ between the RF pulse frequency and the Larmor precession rate $\omega_L$ depends on the local magnetic field at the nuclear spin site, which differs from $B_{ext}$. Hence, the sample comprises $k$ different precession frequencies $\omega_L^{\,k}$ leading to $k$  detunings $\delta_k$. 
 
Our AERIS protocol comprises two parts. The first one creates a detectable signal by exploiting the dynamics  in Eq.~\eqref{bloch}. This is achieved with an RF triggering pulse on the sample followed by an alternation among free precession periods and induced rotations as shown in Fig.~\eqref{fig2}. The second part consists on probing the produced signal with NV sensors that acquire a phase determined by its amplitude, gathering in their spin state information about the spectral composition of the signal and allowing to determine the local magnetic environment around nuclear spins.

More in detail: An RF pulse along the X axis (i.e. $\phi = 0$) of duration $\pi/(2\Omega)$, see Eq.~\eqref{bloch}, tilts the initial thermal polarization of the sample, $\boldsymbol{M}=(0,0,1)$ (note that $\boldsymbol{M}$ provides the direction of the thermal polarization when it is a unit vector, and it attenuates the polarization amount as $T_1$ and $T^*_2$ diminish its modulus) to the perpendicular plane and triggers off the protocol. Once the pulse is turned off, i.e. $\Omega=0$ in Eq.~\eqref{bloch}, the nuclear spins  precess around the external field at a rate determined by the local magnetic field at their respective locations. Similar to a clock mechanism that rotates the needles representing hours, minutes, and seconds at different speeds, the free precession stage of fixed duration $\tau$ splits the magnetization vector $\boldsymbol M(t)$ in components $\boldsymbol M_k (t)= \left( \sin( \delta_k \,t),-\cos( \delta_k t), 0\right)$. Recall, $\delta_k$ is the detuning between the driving frequency and the $k$th precession frequency in the sample. Crucially, the NV sensor remains inactive at this stage, thus $\tau$ could be significantly larger than NV coherence times leading to a high spectral resolution ultimately limited by the coherence of the nuclear sample.

A long RF pulse then continuously rotates the magnetization components at a speed $\propto \Omega$ around an axis on the xy-plane determined by $\phi$, as described by Eq.~\eqref{bloch}. The projection of the resulting field in the NV axis sets a target field  $B(t)$ with two key features. Firstly, it oscillates with frequency $\Omega \ll \omega_L$ (note that, at large magnetic fields, this relation is naturally achieved for realistic Rabi frequencies). This is a parameter that can be tuned such that $B(t)$ oscillations can be tracked by the NV ensemble regardless of the magnetic field value acting on the sample. Secondly, $B(t)$ comprises the radio signals  produced by each rotating $\boldsymbol{M}_k(t) = \left( \sin( \delta_k  \tau), -\cos( \delta_k \tau) \cos(\Omega t), -\cos( \delta_k   \tau) \sin(\Omega t) \right)$, thus it contains the footprint of each nuclear  environment (encoded in the distinct $\delta_k$ shifts). Note that, for the sake of simplicity in the presentation, we do not account for potential deviations in the rotation axes caused by each $\delta_k$ shift. However, these are included in our numerical analysis. For more details see Appendix~\eqref{app1}.

After $N$ complete rotations of the magnetization vectors, thus after N periods of $B(t)$, the RF rotation pulse is switched off and the sample advances to the next free precession stage in which each $\boldsymbol{M}_k(t)$ continues to dephase. This sequence is iterated leading to an oscillation in the amplitudes of the signals emitted during  successive induced rotation stages, whose spectrum relates directly to the various $\omega_L^{\,k}$ in the sample. 

The radio signal $B_n(t)$ produced during the $n^{th}$ induced rotation stage is captured by the NVs in the ensemble such that each NV evolves according to 
\begin{equation}\label{eq:NVHamiltonian}
H/\hbar = - \gamma_e B_n(t) \frac{\sigma_z}{2} + \Omega_{\rm MW}(t) \frac{\sigma_\phi}{2}.
\end{equation}
Here $\gamma_e$ is the electronic gyromagnetic factor, $\boldsymbol{\sigma}$ are the Pauli operators of the NV two-level system, and the target signal $B_n(t)$ is expressed in Appendix~\eqref{app1}.
The control field $\Omega_{\rm MW}(t)$ is synchronized with the rotation pulse over nuclear spins, see Fig.~\eqref{fig2}, leading to an XY4 control sequence that allows the sensor to capture a phase determined by (i) The amplitude of the radio signal stemming from the sample, and (ii) The length of the RF pulse. This information is gathered by reading  the state of the sensor, with an expected result for the $n^{\text{th}}$ phase acquisition stage of 
\beq
\label{expectedsigma}
\langle \sigma_y \rangle_n = \frac{2 \gamma_e t_m }{\pi} \sum_k b_k \cos(\delta_k n \tau),
\eeq
where $b_k$ is the initial magnetic field amplitude on the NV site produced by the $k^\text{th}$ spectral component, see Appendix~\eqref{app1}.
\begin{figure*}[t]
\centering
\includegraphics[width=1.\linewidth]{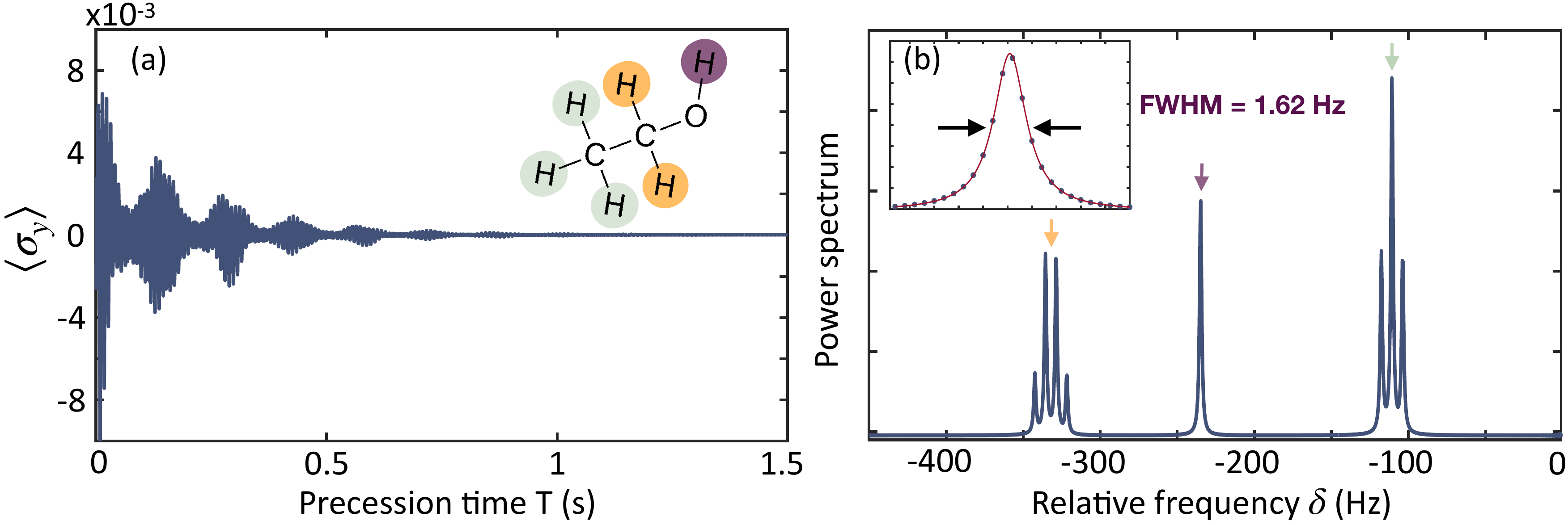}
\caption{\label{fig:results}  Measurements and spectrum obtained by considering $\delta_k = -\{342.45, 335.55, 328.65, 321.75, 234.9, 117.6, 110.7, 103.8\}$ Hz, and magnetic field amplitudes $b_k = \{106 ,320, 320, 106, 426, 320, 640, 320\}$ pT along the Z axis of a generic NV in the ensemble. (a) Simulated stroboscopic record collected by measuring $\langle \sigma_y \rangle$ on the NV as a function of the cumulated precession time, after interacting with the ethanol sample (inset). The three sites of the ethanol molecule with different chemical shifts are indicated with distinct colors. (b) Fourier transform of the measurement record (blue solid line) showing peaks in the expected values. Each peak group has its origin site/chemical shift indicated with an arrow of the corresponding color. Inset, the central peak was fitted to a Lorentzian function that exhibits a full width at half maximum (FWHM) of 1.62 Hz.}
\end{figure*}
Thus, subsequent detections provide a stroboscopic record of the oscillating amplitudes, see Fig.~\eqref{fig:results} (a), whose Fourier spectrum relates to the frequency shifts of nuclei at different sites of the sample molecule.

Let us recall that the NV ensemble sensor is only active during phase acquisition (i.e. while the dynamical decoupling sequence is active), and after that, it is optically readout and reinitialized. Therefore, the duration of our  protocol, and thus its spectral resolution, gets over the cap imposed by the coherence of the sensor, being only limited by the coherence of the nuclear fields.

\emph{Numerical Results.-} We illustrate the AERIS protocol by simulating the evolution of 8 magnetization vectors taken from the ethanol [C$_2$H$_6$O] spectrum~\cite{Levitt08} in a scenario that comprises a magnetic field of 2.1 T, while the RF driving frequency $\omega$ is set to $(2\pi)\ \times$ 90 MHz, which is assumed to be the origin of the chemical shift scale (this is the resonance frequency of TMS~\cite{Levitt08}). Each $\delta_k$ detuning is obtained by considering the three chemical shifts of $3.66$, $2.6$, and $1.19$ ppm, as well as a J-coupling of 6.9 Hz between the CH$_3$ and the CH$_2$ groups of ethanol~\cite{Levitt08}, see caption in Fig.~\eqref{fig:results}. The average field amplitude over each NV in the ensemble is estimated to  $\approx2.56$ nT, by taking into account the proton concentration of ethanol as well as the external magnetic field of 2.1 T, see Appendix~\ref{app2}. This  field amplitude is distributed in different $b_k$ according to the ethanol spectral structure, see caption in Fig.~\eqref{fig:results} and Appendix~\ref{app2}. We find the radio signal emitted by the sample by numerically solving the Bloch equations during RF irradiation (i.e. at the induced rotation stages). The free precession time is selected as $\tau = 1$ ms, and the induced rotation stage has a duration of 40 $\mu$s (corresponding approximately to 2 full rotations of the magnetization vectors) while the NV ensemble is controlled with an XY4 sequence. Furthermore, we use $\Omega_{\rm MW} = (2 \pi)\times 20$ MHz, $\Omega_{\rm RF} = (2 \pi)\times 50$ KHz, and sample coherence times  $T_1 = 2$ s and $T^*_2 = 0.2$ s. This process is repeated 1500 times, leading to the stroboscopic record of Fig.~\eqref{fig:results} (a) which follows Eq.~(\ref{expectedsigma}). 

We run again the protocol by employing an initial $\pi/2$ pulse over the Y axis leading to the sinusoidal version of  Eq.~(\ref{expectedsigma}). This is:
\beq\label{expectedsigma2}
\langle \sigma_y \rangle_n = \frac{2 \gamma_e t_m }{\pi} \sum_k b_k \sin(\delta_k n \tau).
\eeq
Finally, both measurement records in Eqs.~(\ref{expectedsigma}, \ref{expectedsigma2}) are combined and converted, via discrete Fourier transform, into the spectrum in Fig.~\eqref{fig:results} (b). There we demonstrate that the AERIS method leads in the studied case to Lorentzian peaks with a FWHM  $\approx 1.62$ Hz (limited by the sample $T^*_2$) thus sufficient to detect the posed chemical shifts and J couplings.

For the sake of simplicity in the description of the AERIS method, the presented simulations consider perfect controls. Appendix. \eqref{app3} analyses the impact of faulty RF driving. We find that, for realistic errors \cite{Boris22, Cai12}, the method still provides results that resemble the ideal ones. Moreover, for more severe error levels, in Appendix. \eqref{app3} we devise an alternative AERIS sequence that enhances the robustness of the protocol.

\emph{Conclusions.-} We have devised an NMR signal detection protocol that attains chemical shift level resolution from micron-sized samples while being suitable for large magnetic fields. Our approach relies on the production of a custom field that resonates with dynamically decoupled NV sensors used to extract spectral information from the sample. Actual experiments may require several repetitions to average out the impact of shot noise or inaccurate control sequences. Nevertheless the demand for higher spectral resolution is less stringent at large fields, as chemical shifts increase and J-couplings become clearer. Besides, polarization rates increase, leading to stronger signals that provide measurements with higher contrast.  Both effects contribute to decreasing the required number of repetitions, or, conversely, making small concentration samples amenable to our protocol, which sets the utility of NV sensors for realistic chemical analysis.

% Finally, our method could accommodate distinct upgrades, such as (i) Pre-polarization techniques that provide stronger and coherent nuclear signals even from nanoscale samples, (ii) More sophisticated dynamical decoupling techniques that augment the resilience of the operation to noise sources, and (iii) Additional RF drivings over the sample (such as Lee Goldburg \cite{Lee65} or WAHUHA schemes \cite{Waugh68}) that would  further contribute to remove dipolar interactions leading to cleaner spectra.

\begin{acknowledgements}
\emph{Acknowledgements.--}
We thank fruitful discussions with A. Mart\'in. C.M.-J. acknowledges the predoctoral MICINN grant PRE2019-088519. J.~C. acknowledges the Ram\'{o}n y Cajal   (RYC2018-025197-I) research fellowship, the financial support from Spanish Government via EUR2020-112117 and Nanoscale NMR and complex systems (PID2021-126694NB-C21) projects, the EU FET Open Grant Quromorphic (828826),  the ELKARTEK project Dispositivos en Tecnolog\'i{a}s Cu\'{a}nticas (KK-2022/00062), and the Basque Government grant IT1470-22.

\emph{Note added.-} In the preparation of the manuscript, we become aware of a similar concept using a double electron-nuclear resonance to detect NMR spectra \cite{Meinel22}. 
\end{acknowledgements}

\onecolumngrid
\appendix

\section{Measured signal\label{app1}}

After the triggering pulse with $\phi=0$, the magnetization reads ${\boldsymbol M}=(0,-1,0)$ . This is the initial configuration for the first free precession stage (of fixed duration $\tau$) that splits the magnetization in different components ${\boldsymbol M}_k$ such that the initial configuration for the $k^\text{th}$ spectral component at the $n^\text{th}$ phase acquisition stage reads ${\boldsymbol M} = (\sin(\delta_k n \tau), -\cos(\delta_k n \tau),0)$. This is obtained from the Bloch equations with $\Omega=0$. 

Now, the evolution of the magnetization during the $n^{th}$ phase acquisition stage reads 
\beq
\label{solbloch}
\boldsymbol{M}_k^{\,n}(t)=
\left(\begin{array}{ccc}
1 & 0 & 0\\
0 & \cos(\Omega t) & - \sin(\Omega t)\\
0  & \sin(\Omega t) & \cos(\Omega t)
\end{array}\right)
\left(\begin{array}{c}
\sin( \delta_k \,n \tau)\\
-\cos( \delta_k \, n \tau)\\
0
\end{array}\right)
=
\left(\begin{array}{c}
\sin( \delta_k \,n \tau)\\
-\cos( \delta_k \, n \tau) \cos(\Omega t)\\
-\cos( \delta_k \, n \tau) \sin(\Omega t)
\end{array}\right).
\eeq 
Notice that the Bloch equations in the main text, and therefore the solutions obtained from them, describe the dynamics in a frame that rotates around the external field at the RF driving frequency $\omega$. For the sake of simplicity in the presentation, the solution in Eq.~\eqref{solbloch} assumes no decoherence (i.e. $T_1,T^*_2 \rightarrow \infty$) and perfect resonance (not taking into account the natural $\delta_k$ shifts during the driving. However, the numerical simulations leading to the results displayed in the main text include realistic $T_1$ and $T^*_2$, as well as the corresponding $\delta_k$ shifts, and hence imply numerically solving the Bloch equations in Eq.~\eqref{bloch} of the main text. Here we show the approximate analytical solution to provide the reader with an insight into the dynamics at the phase acquisition stages of the protocol. 

In our case, the effect of the $\delta_k$ shifts on the rotation speed is, to first order, a factor of approximately $\frac{\delta_k^2}{2 \Omega^2} \approx 2 \times 10^{-5}$, which is negligible and has no significant impact on the results. If necessary (i.e., in case of facing more severe energy shifts) a modified sequence, as outlined in Appendix~\eqref{app3}, can be used to further correct this error.

The interaction between the signal produced by the rotating $\boldsymbol{M}_k^{\,n}(t)$ and the sensor in a rotating frame w.r.t. the NV electronic-spin ground-state triplet is 
\beq
H/\hbar = - \gamma_e B_n(t) \frac{\sigma_z}{2} + \Omega_{\rm MW}(t) \frac{\sigma_\phi}{2}.
\eeq
Here $\Omega_{\rm MW}(t)$ is the MW control field, and the target signal induced by the $n^{th}$ phase acquisition stage is $B_n(t)=\sum_k B_k^n(t)$ such that
\beq
\label{targetfieldapp}
B_k^n(t) = \frac{\hbar^2 \gamma_N^2 \mu_0 \rho_k  B_{ext}}{16 \pi k_B T}\, {\boldsymbol M}_k^{\,n}(t) \int  \big[g_x(r), g_y(r), f(r)\big] \ dV,
\eeq
where $\mu_0$ is the vacuum permeability, $\rho_k$ the density of spins with the $k^{th}$ precession frequency, $\gamma_N$ is the nuclear gyromagnetic factor, $T$ is the temperature of the sample, $k_B$ is the Boltzmann constant, and $B_{ext}$ is the external magnetic field. The geometric functions $g_{x,y}(r)$ and $f(r)$ read
\beq
f(r)= \frac{1}{r^3}(3 r_z^2 -1) \hspace{1.5cm} \text{and} \hspace{1.5cm} g_{x, y}(r)= \frac{1}{r^3} (3 r_z r_{x, y}),
\eeq
with $\hat{r} = (r_x, r_y, r_z)$ being  the unitary vector joining the NV and $dV$, while $r$ represents their relative distance. The expression in~\eqref{targetfieldapp} (which can be derived from a microscopic description of a system involving NVs and nuclear spins \cite{Meriles10}) is valid provided that the external magnetic field $B_{ext}$ is greater than the coupling strength, which allows ignore the backaction of the sensor in the sample \cite{Reinhard2012}. As we are in a large field regime, this condition is met. In addition, the contribution of the orthogonal components $M^{\,n}_{k,x}(t)$ and $M^{\,n}_{k,y}(t)$ to $B_k^n(t) $ (which rapidly oscillate with the Larmor frequency $\gamma_N B_{ext}$) can be safely neglected.

The MW control implements an XY4 dynamical decoupling sequence that modulates the interaction between target and sensor leading to 
\beq
H/\hbar = \frac{\gamma_e  \sigma_z}{\pi}  \sum_k b_k \cos(\delta_k n \tau),
\eeq
where $b_k = \frac{\hbar^2 \gamma_N^2 \mu_0 \rho_k  B_{ext}}{16 \pi k_B T}  \int  f(r) dV $. 

 The NV is initialized in the $|+\rangle = \frac{1}{\sqrt{2}}\left(|1\rangle+|0\rangle\right)$ state, then evolves during $t_m$, and it is finally measured such that (in the small angle regime)
\beq
\langle \sigma_y \rangle_n = \frac{2 \gamma_e t_m}{\pi}  \sum_k b_k \cos(\delta_k n \tau).
\eeq

On the other hand, an RF trigger pulse with $\phi=\pi/2$ leads to $\boldsymbol{M}_k=(1,0,0)$, which yields a splitting of the $k$ spectral components during the free precession stages described by  ${\boldsymbol M}_k = (\cos(\delta_k n \tau),\sin(\delta_k n \tau),0)$. For the same dynamical decoupling control sequence over NVs, we find   

 \beq
\langle \sigma_y \rangle_n = \frac{2 \gamma_e t_m }{\pi} \sum_k b_k \sin(\delta_k n \tau).
 \eeq

\section{Radio field intensity estimation\label{app2}}

In this section, we estimate the radio signal amplitude for the example in the main text. We numerically compute the geometrical integral $F =  \int  f(r) dV$ for different hemispheres while we consider the NV axis perpendicular to the diamond surface. This leads to an asymptotical value of $F \sim4.1$. Note that half of the asymptotic value is reached for integration hemispheres with a radius of 2-3 times the depth of the NV, which leads to detectable signals even for picoliter volume samples. Considering a pure ethanol sample with a density of 789 kg m$^{-3}$ and a molar mass of 46 g mol$^{-1}$, we obtain a proton density of $\rho =$ 6.2 $\times\ 10^{28}$ m$^{-3}$.  With this into consideration, the total amplitude obtained in a 2.1 T external field at room temperature is $b\sim 2.56$ nT. Finally, we can distribute this amplitude throughout the ethanol spectral peaks according to the following rules: $b/3$ (signal produced by 2 out of 6 hydrogens of the molecule) distributed in four peaks with ratios 1:3:3:1, a single peak of $b/6$, and $b/2$ (signal produced by 3 out of 6 hydrogens of the molecule) distributed in three peaks with ratio 1:2:1, to obtain
\beq
b_k \in \{106 ,320, 320, 106, 426, 320, 640, 320\}\ {\rm pT}.
\eeq

\section{Sequence robustness considerations\label{app3}}

\begin{figure*}[h]
\centering
\includegraphics[width=0.7\linewidth]{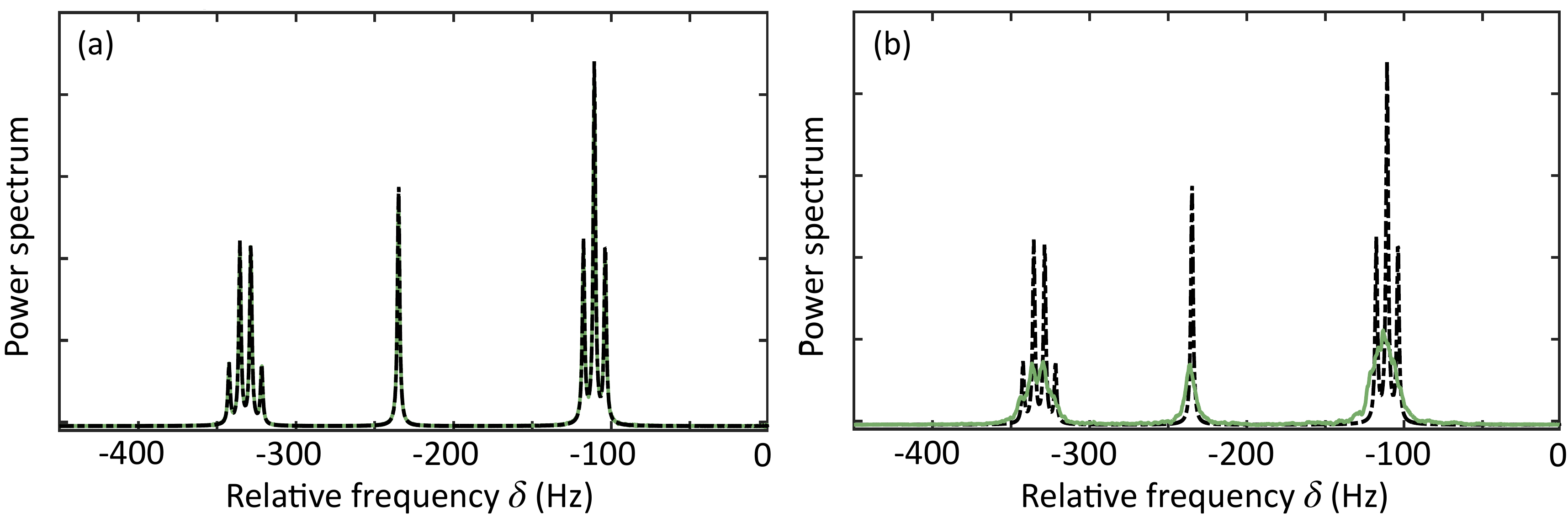}
\caption{\label{fig:appRobust1}  Spectra comparison for the AERIS sequence with perfect RF controls (black dotted line) and in the presence of OU noise (green solid line) with (a) $\sigma = 0.24\%$ and $\tau = 1$ms and (b) $\sigma = 2\%$, $\tau = 0.5$ms and an amplitude shift of 1$\%$. Both spectra were obtained averaging 200 realizations.}
\end{figure*}

We consider the effect of errors in the RF control, which could be potentially detrimental for the sequence as the nuclear signal coherence has to be maintained throughout the protocol. The control error is modeled as an Ornstein-Uhlenbeck \cite{Wang45, Gillespie96} process
\beq
\epsilon_\Omega(t+\Delta t) = \epsilon_\Omega(t) e^{-\Delta t/\tau} + \sigma N(t)
\eeq
where $\tau$ is the correlation time of the noise, $N(t)$ is a normally distributed random variable, and $\sigma$ is the relative amplitude of the fluctuations. For standard expected experimental errors \cite{Boris22, Cai12}, the obtained spectrum overlaps with the case without control errors, see Fig.~\ref{fig:appRobust1} (a).

\begin{figure*}[h]
\centering
\includegraphics[width=0.9\linewidth]{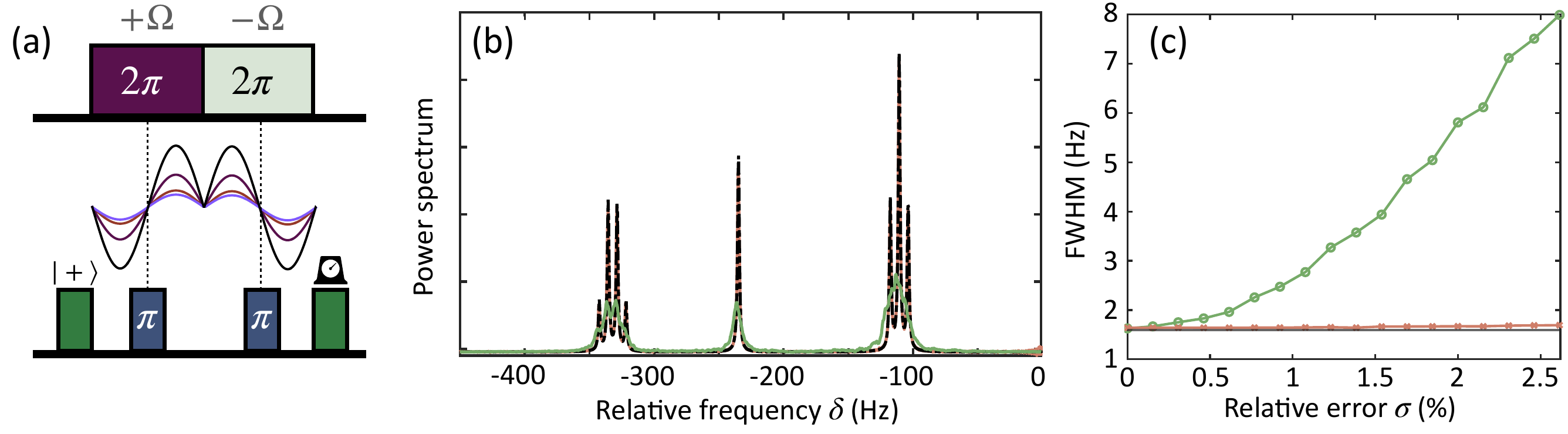}
\caption{\label{fig:appRobust2}  (a) Schematics of the modified AERIS sequence. Two rotations are performed in opposite directions with the RF control, giving raise to a detectable magnetic signal with a $\pi$ phase change in the middle which is measured with two concatenated spin echoes. (b) Spectra comparison for the AERIS sequence with perfect controls (black dotted line) and with $\sigma = 2\%$, $\tau = 0.5$ms and an amplitude shift of 1$\%$ for AERIS (green solid line) and the modified version (orange solid line). (c) FWHM with respect of the relative OU error with $\tau = 1$ms and no constant amplitude shift for AERIS (green line) and the modified version (orange line). The minimum FWHM possible given the nuclear $T_2^*$ is represented as a grey line.}
\end{figure*}

However, in the presence of more severe noise and constant Rabi amplitude shifts (e.g. due to misscalibration) AERIS gives raise to distorted spectra as can be seen in Fig.~\ref{fig:appRobust1} (b). A direct modification of the default sequence leads to a significant improvement on robustness. The alternative sequence is equivalent to the original one but changes the irradiation/NV measurement stages with the scheme represented in Fig.~\ref{fig:appRobust2} (a). The modified version employs a change of sign in the middle of the RF irradiation such that the error accumulated in the first half is the opposite to the one accumulated in the second half leading to cancellation. The XY-4 sequence over the NV is substituted with two $\pi$ pulses in order to accumulate phase from the new magnetic signal. The new version recovers the ideal spectrum in the severe noise example, see Fig.~\ref{fig:appRobust2} (b).

Finally, in Fig.~\ref{fig:appRobust2} (c), we show a comparison of the expected FWHM of the central spectral peak for AERIS and the modified version with respect to the error amplitude. The modified version recovers a FWHM close to the minimum possible given the nuclear $T_2^*$ for the considered error range.

\end{document}